\documentclass{ws-ijmpc}

\begin{document}

\markboth{Nuno Crokidakis}
{Radicalization phenomena: Phase transitions, extinction processes and control of violent activities}

\catchline{}{}{}{}{}
%%%%%%%%%%%%%%%%%%%%%%%%%%%%%%%%%%%%%%%%%%%%%%%%%%%%%%%%%%%%%%%%%%%%

\title{Radicalization phenomena: Phase transitions, extinction processes and control of violent activities}

\author{Nuno Crokidakis $^{*}$}

\address{
Instituto de F\'{\i}sica, \hspace{1mm} Universidade Federal Fluminense \\
 Niter\'oi - Rio de Janeiro, \hspace{1mm} Brazil \\ 
$^{*}$ nunocrokidakis@id.uff.br}

\maketitle

\begin{history}
\received{Day Month Year}
\revised{Day Month Year}
\end{history}

\begin{abstract}
\noindent
In this work we study a simple mathematical model to analyze the emergence and control of radicalization phenomena. The population consisits of core and sensitive subpopulations, and their ways of life may be at least partially incompatible. In such a case, if a conflict exist, core agents act as inflexible individuals about the issue. On the other hand, the sensitive agents choose between two options: live peacefully with core population, or oppose it. This kind of modeling was recently considered by Galam and Javarone (2016) with constant pairwise couplings. Here, we consider the more general case with time-dependent transition rates, with the aim of study the impact of such time dependence on the critical behavior of the model. The analytical and numerical results show that the nonequilibrium active-absorbing phase transition can be suppressed in some cases, with the destruction of the absorbing phase where the radical agents disappear of the population in the stationary states.


\end{abstract}

\ccode{PACS Nos.: 05.10.-a, 05.70.Jk, 87.23.Ge, 89.75.Fb}

\section{Introduction}

\qquad Radicalization phenomena occur in many countries. As a collective social phenomena, it can be studied in the context of statistical physics of complex systems \cite{social_rmp}. Indeed, analytical and numerical tools of statistical physics were considered in the study of several social dynamics like political polarization \cite{nuno_galam,kaufman,nuno_moderate}, opinion dynamics \cite{sznajd,galam_review}, adoption of innovations \cite{iglesias2}, corruption \cite{rafael_nuno}, rumor spreading \cite{liqing}, alcoholism \cite{nuno_lucas1,nuno_lucas2,jin}, criminality \cite{nuno_applied,iglesias}, and many others.

The term radicalization refers to the process of developing extremist religious political or social beliefs and ideologies. It is a process through which people become increasingly motivated to use violent means against members of an out-group or symbolic targets to achieve behavioral change and political goals. An important feature of radical groups is that most of such groups embrace an ideology that legitimizes violence to address their concerns, and this violence is often directed at an out-group viewed as the culprit responsible for creating the grievance \cite{doosje}. While radical thinking is by no means problematic in itself, it becomes a threat to national security when it leads to violence \cite{mcCluskey}.  

Recent data from the European Council of the European Union \cite{eu_data} indicated that EU countries reported: (1) a total of 15 completed, failed and foiled terrorist attacks in 2021; (2) Jihad terrorists completed 3 attacks in France, Germany and Spain; (3) 2 people were killed in terrorist attacks in the EU in 2021 (Jihadi terrorists were responsible for both deaths); (4) 388 arrests on suspicion of terrorist offences in EU countries were reported to Europol in 2021; (5) In 2021, the most frequent offence leading to arrest - among those reported - was membership of a terrorist organisation, often combined with propaganda dissemination or planning terrorist acts. Thus, the understanding of such social phenomena is of interest for several areas of science.

To deal with radicalization phenomena, some mathematical models were proposed \cite{santoprete,santoprete2,nathan,wang,javarone_galam}. Santoprete considered the strategy known as Countering Violent Extremism, taking into account prevention and de-radicalization\cite{santoprete}. The global stability of the model was studied, and strategies to counter violent extremism were discussed. The model was extended to consider a vaccination compartment, in order to describe individuals in prevention programs. The results suggest that de-radicalization seems to be more effective to counter radicalization than prevention \cite{santoprete2}. Another mathematical model was introduced in order to to describe prevention programs in marginalized population by incorporating government inclusivity. Numerical simulation of the model carried out showed that enhanced government inclusivity leads to a slower rate of transition to radical population \cite{nathan}. Considering multiple ideologies, a recent work showed show that ideologies with cooperative mechanisms are easier to establish themselves in a group and are difficult to eliminate. This makes it more difficult to curb radicalization of the population \cite{wang}. The phenomenon of radicalization was also investigated within a mixed population composed by core and sensitive subpopulations. The results highlight the instrumental role core agents can have to hinder radicalization within the sensitive subpopulation \cite{javarone_galam}. We can also highligth the results of Ref. \cite{javarone_galam2}, where the authors studied the role os passive supporters on the spreading of extremist opinions. The model is 3-state opinion model, where the three opinions represent pro-western opinion, anti-western opinion and extreme anti-western opinion. They discussed that a substantial fraction of anti-western agents adopt the extreme opinion exhibiting an emergent phenomenon which may shed some new light on real social phenomena of political violence  \cite{javarone_galam2}.

Compartmental models have been used recently to deal with a plethora of social contagion phenomena (for a recent review, see \cite{sooknanan}). In this work, we follow this approach to study the phenomena of radicalization. For this purpose, we follow \cite{javarone_galam} and consider a population consisting of core and sensitive subpopulations. The system evolves under pairwise interactions among agents, and we consider time-dependent transition rates, with the aim of study the impact of such time dependence on the critical behavior of the model. The analytical and numerical results show that the population undergoes nonequilibrium active-absorbing phase transitions, that can be suppressed in some cases, with the destruction of the absorbing phase where the radical agents disappear of the population in the stationary states. 

This work is organized as follows. In section 2 we introduce the mathematical formulation of the model. In section 3 we introduce the time-dependent rates, considering two distinct cases. Finally, in section 4 we present our final remarks.

% ###########################################################################

\section{Model}

\qquad We consider a heterogeneous population composed by sensitive and core agents. Following \cite{javarone_galam}, the sensitive population is composed by opponents and peaceful agents, which densities are denoted by $\sigma_O(t)$ and $\sigma_P(t)$, respectively, at a given time step $t$. The social dynamics is ruled by pairwise interactions. These interactions occur both within the sensitive population and by mixing with core agents.  In case of a conflict, such core agents behave as inflexible about the issue \cite{javarone_galam}, considering a given fixed density $\sigma_I$ of inflexibles. We will consider pairwise interactions. For such purpose, we introduce a parameter $\alpha(t)$ denoting the rate per unit of time of encounters where opponents become peaceful agents. In addition, we consider another parameter $\beta(t)$ to account for the rate of success of opponents in convincing peaceful agents to turn opponents. The possible transitions are given squematically as:
\begin{eqnarray} \label{eq1}
\sigma_O + \sigma_I  \stackrel{\alpha(t)}{\rightarrow} & \sigma_P + \sigma_I ~,  \\ \label{eq2}
\sigma_P + \sigma_O  \stackrel{\beta(t)}{\rightarrow} & \sigma_O + \sigma_O ~.
\end{eqnarray}
Considering a fully-connected population, we can write the rate equations based on the possible transitions given by Eqs. \eqref{eq1} and \eqref{eq2}. Thus, the following system of differential equations governs the dynamics of the population:
\begin{eqnarray} \label{eq3}
\frac{d}{dt}\,\sigma_P(t) & = & \alpha(t)\,\sigma_I\sigma_O(t) - \beta(t)\,\sigma_O(t)\,\sigma_P(t) \\  \label{eq4}
\frac{d}{dt}\,\sigma_O(t) & = & \beta(t)\,\sigma_O(t)\,\sigma_P(t) - \alpha(t)\,\sigma_I\sigma_O(t) 
\end{eqnarray}
\noindent
together with the normalization condition, namely
\begin{equation}\label{eq5}
\sigma_P(t) + \sigma_O(t) + \sigma_I = 1 ~,
\end{equation}
\noindent
since the population is fixed.

The work \cite{javarone_galam} considered constant rates $\alpha(t)=\alpha$ and $\beta(t)=\beta$. Despite the fact that the authors in Ref. \cite{javarone_galam} did not discuss the phase transition in details, we can see from Ref. \cite{javarone_galam} that, for $\alpha$ and $\beta$ constant, the population undergoes an active-absorbing nonequilibrium phase transition \cite{dickman_book} at critical point $\beta_c=\alpha\sigma_I/(1-\sigma_I)$. For such transition, we can consider the stationary fraction of opponents, $\sigma_O$, as the order parameter. For $\beta\leq\beta_c$, the opponents disappear of the population in the stationary states, and the population is composed only by inflexible and peaceful agents. On the other hand, for $\beta>\beta_c$ the radicalization persists in the long-time limit, and the three subpopulations coexist. Our target in this work is to study the impact of time-dependent transitions rates $\alpha(t)$ and $\beta(t)$ on the dynamics of the model, as well as in the occurrence of phase transitions.

In the next sections we consider two distinct cases, according to the time-dependence of the transitions rates: (i) $\alpha$ constant and $\beta=\beta(t)$ and (ii) $\beta$ constant and $\alpha=\alpha(t)$.

% ############################################################################

\section{Results}

\subsection{Case 1: $\alpha=$ constant, $\beta=\beta(t)$}

\qquad In this section we consider a kind of feedback mechanism in the social interaction ruled by $\beta$. In other words, since $\beta$ is a measure of the persuasive power of opponents over peaceful agents, we considered that $\beta$ is a function of time that depends on the density of peaceful individuals $\sigma_P(t)$. Thus, given that the density of peaceful agents vary with time, the opponents adjust their ability to convice peaceful individuals to become opponents. In other words, we considered the following form for $\beta(t)$,
\begin{equation}\label{eq6}
\beta(t) = \beta_0 + \beta_1\,\sigma_P(t) ~.
\end{equation}

As discussed in \cite{javarone_galam}, opponent agents are activists. Thus, from Eq. \eqref{eq6} we can see that if the density $\sigma_P(t)$ increases, the opponents react in order to offset the decrease of opponent agents and increase their ability to convince peaceful agents to become opponents. Thus, the system of Eqs. \eqref{eq3} and \eqref{eq4} becomes
\begin{eqnarray} \label{eq7}
\frac{d}{dt}\,\sigma_P(t) & = & \alpha\,\sigma_I\sigma_O(t) - \beta_0\,\sigma_O(t)\,\sigma_P(t) - \beta_1\,\sigma_O(t)\,\sigma_P^{2}(t)  \\  \label{eq8}
\frac{d}{dt}\,\sigma_O(t) & = & \beta_0\,\sigma_O(t)\,\sigma_P(t) + \beta_1\,\sigma_O(t)\,\sigma_P^{2}(t) - \alpha\,\sigma_I\sigma_O(t)
\end{eqnarray}

Thus, we can see from above Eqs. \eqref{eq7} and \eqref{eq8} that the form given by Eq. \eqref{eq6} introduces another nonlinearity in the model (see the term $\sigma_P^{2}(t)$).

\begin{figure}[t]
\begin{center}
\vspace{6mm}
\includegraphics[width=0.4\textwidth,angle=0]{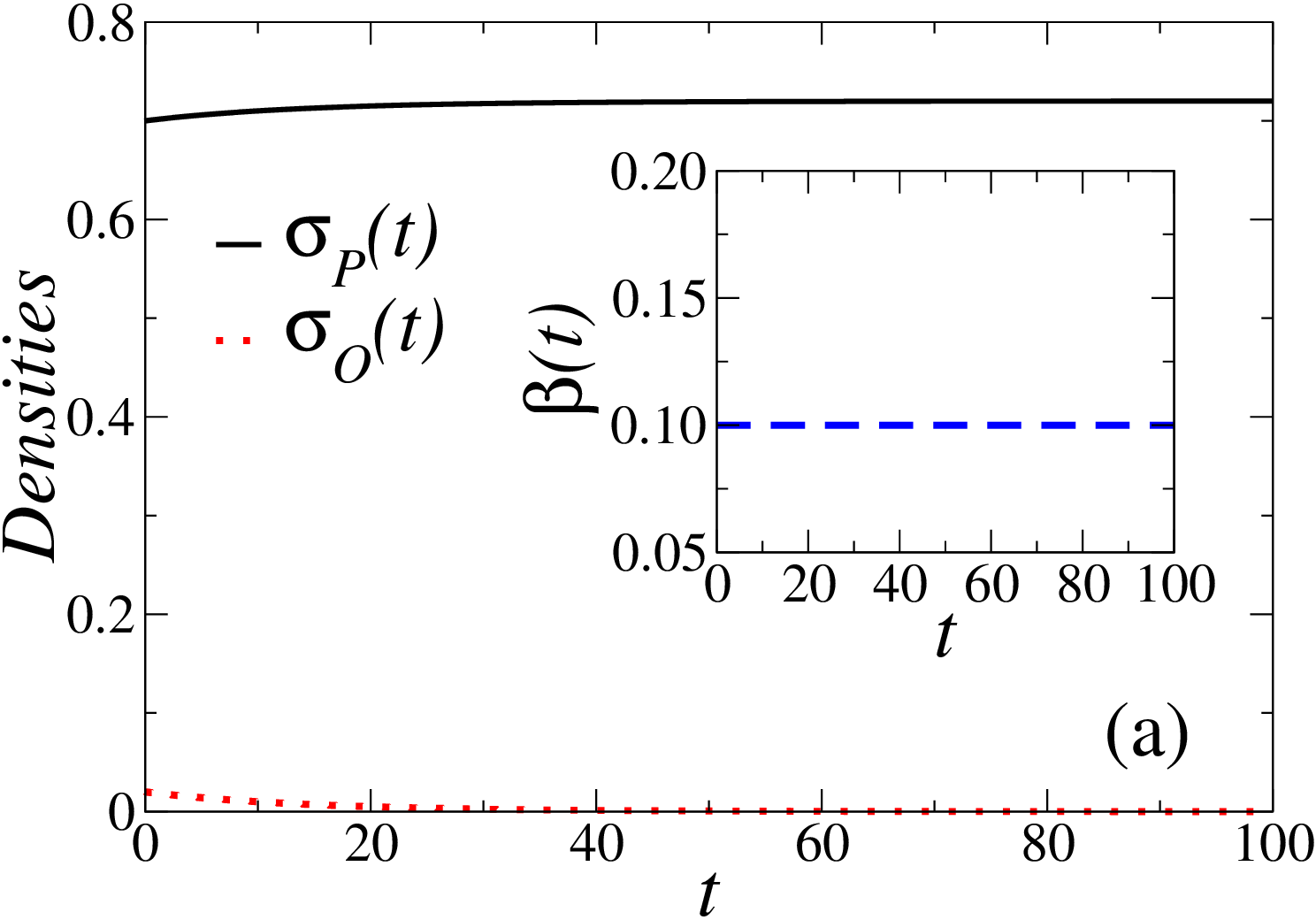}
\hspace{0.2cm}
\includegraphics[width=0.4\textwidth,angle=0]{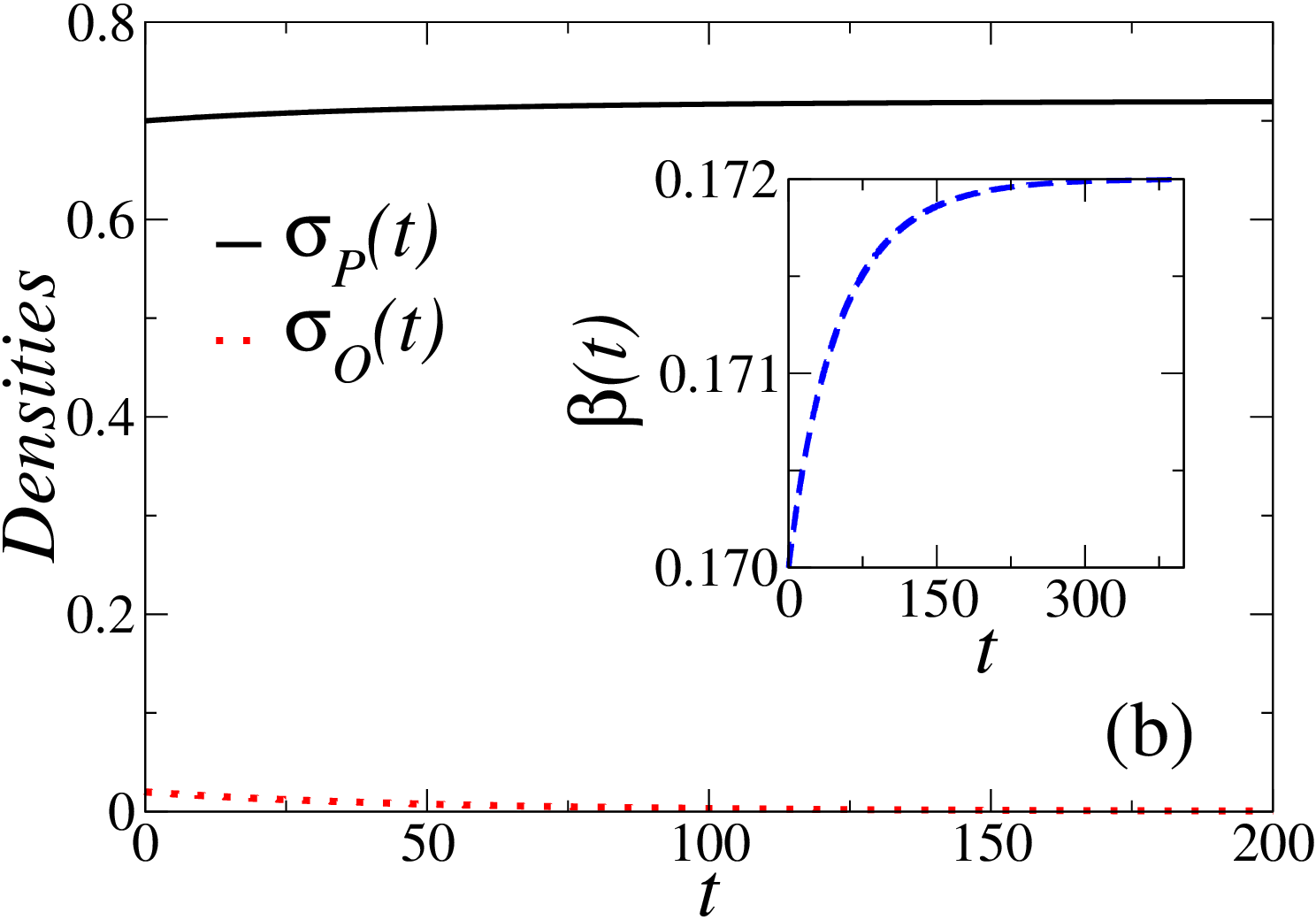}
\\
\vspace{1.0cm}
\includegraphics[width=0.4\textwidth,angle=0]{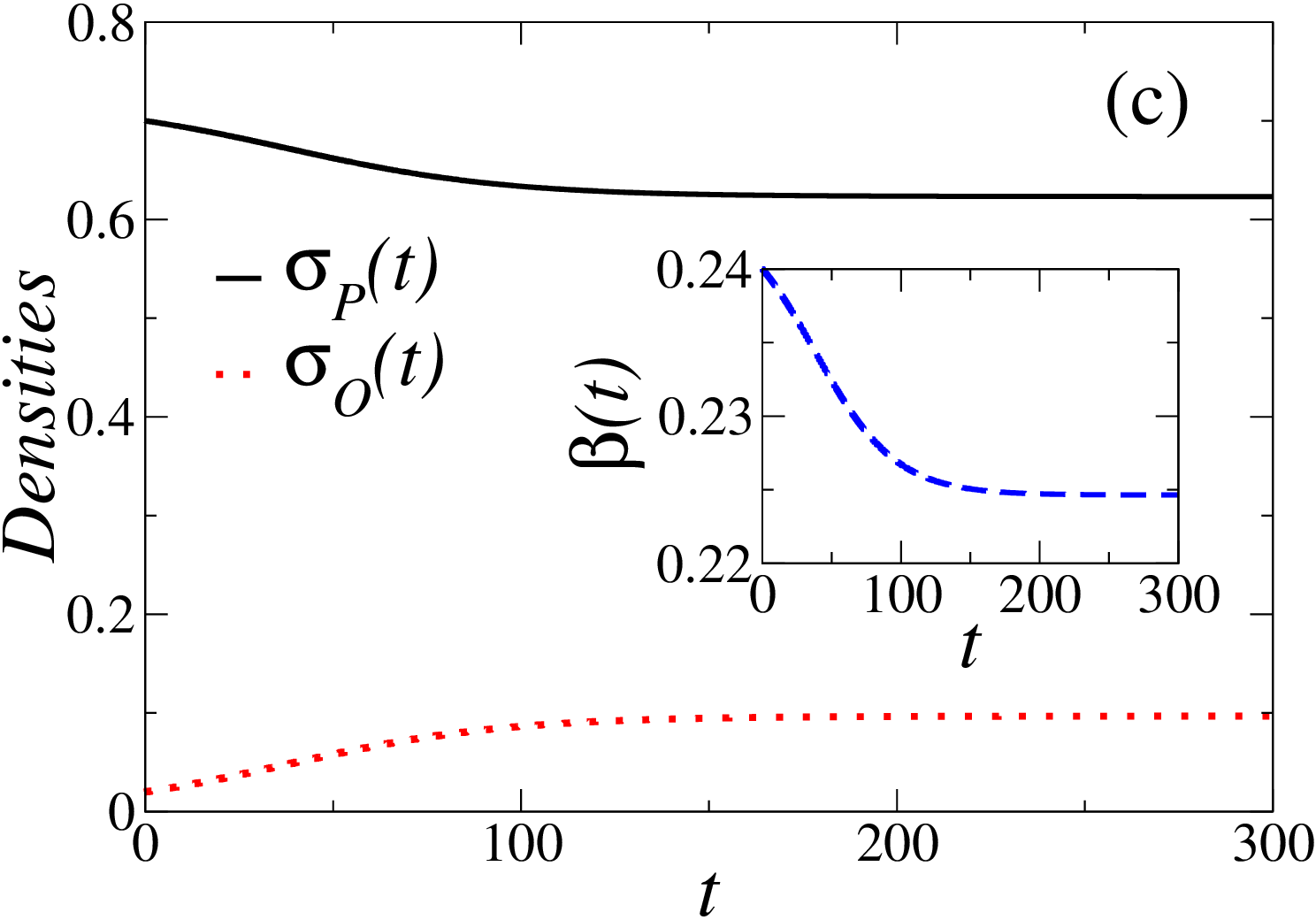}
\hspace{0.2cm}
\includegraphics[width=0.4\textwidth,angle=0]{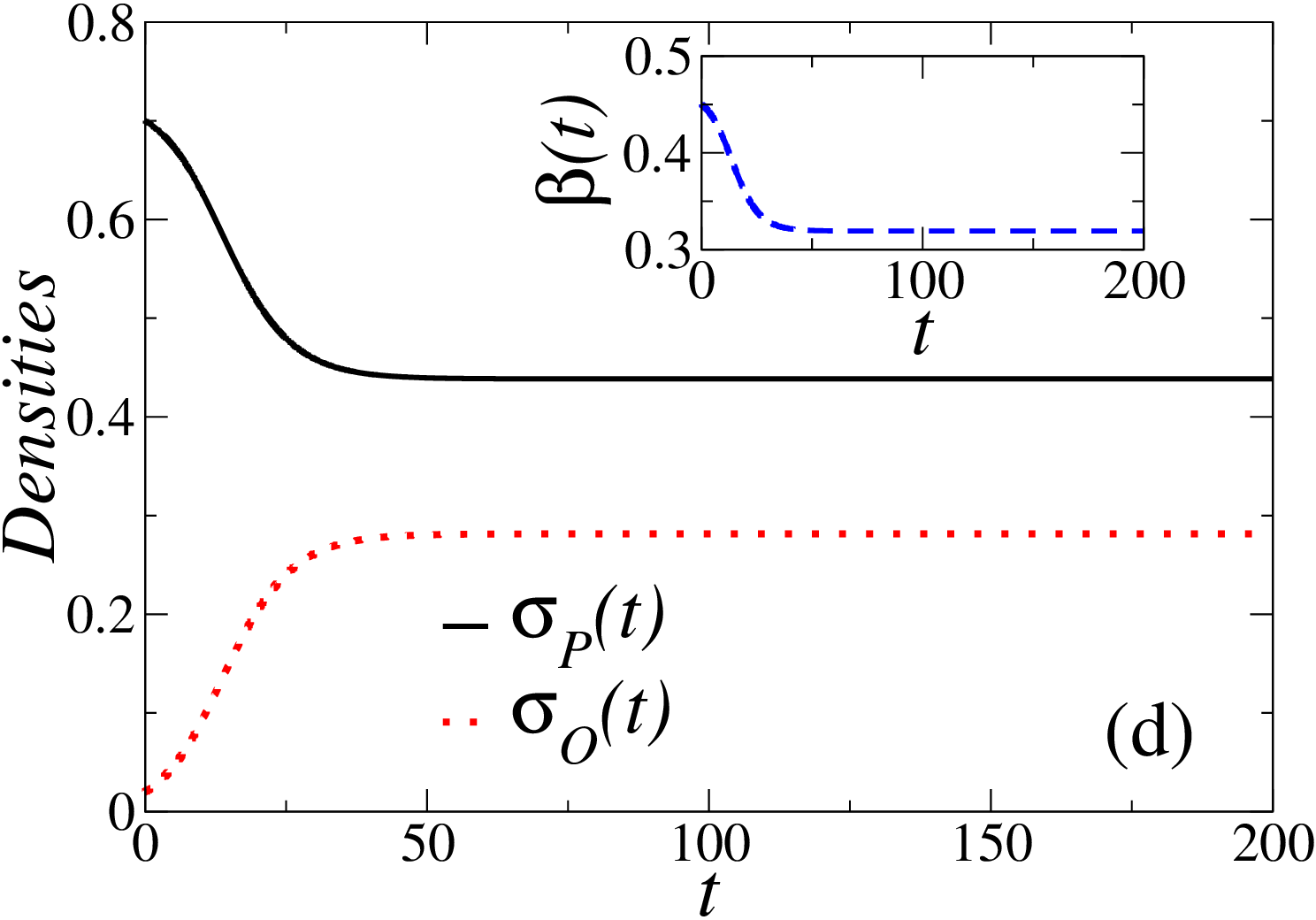}
\end{center}
\caption{(Color online) Time evolution of the densities of peaceful agents $\sigma_P(t)$ and opponents $\sigma_O(t)$ for typical values of $\beta_1$ for the model with time-dependent rate $\beta=\beta(t)$: (a) $\beta_1=0.0$ (b) $\beta_1=0.1$ (c) $\beta_1=0.2$; (d) $\beta_1=0.5$. The insets exhibit the time evolution of the rate $\beta(t)$. The fixed parameters are $\alpha=0.5, \sigma_I=0.28$ and $\beta_0=0.1$.}
\label{fig1}
\end{figure}

We can start analyzing the time evolution of the densities $\sigma_P(t)$ and $\sigma_O(t)$. In Fig. \ref{fig1} we exhibit these quantities for fixed $\alpha=0.5$, $\sigma_I=0.28$ and $\beta_0=0.5$, a typical situation analyzed in Ref. \cite{javarone_galam}. The initial condition is given by  $\sigma_O(0)=0.02$ and $\sigma_P(0)=1- \sigma_O(0) - \sigma_I = 0.70$. The curves were obtained by the numerical integration of Eqs. \eqref{eq7} and \eqref{eq8}. Since the parameter $\beta_1$ is the novelty of the present model, we present graphics in Fig. \ref{fig1} for typical values of $\beta_1$. We can see that the quantities evolve to stationary states after a long time. For $\beta_1=0$ (panel a) we have $\beta(t)=$ constant, and the stationary states are given by $\sigma_O=0$, which represents an absorbing state since for $\sigma_O=0$ we do not have transitins between the states $\sigma_O$ and $\sigma_P$ (see Eqs. \eqref{eq1} and \eqref{eq2}). The case $\beta_1=0.1$ (panel b) also shows this absorbing state for long times, even in a case where the rate $\beta(t)$ evolves in time, increasing its value until stabilizes after some time (see the inset in panel b). Thus, such increase in the rate $\beta$ is not sufficient to change the macroscopic state of the population at stationary states. On the other hand, for the cases $\beta_1=0.2$ (panel c) and $\beta_1=0.5$ (panel d) the rate $\beta(t)$ starts with a higher value in comparison with the previous cases, but it decreases in time since $\sigma_P(t)$ decreases. Even with such decrease in $\beta(t)$ the system evolves to stationary states where the subpopulations of peaceful agents $\sigma_P$ and opponents $\sigma_O$ coexist in the population with the inflexible individuals $\sigma_I$. These results will be discussed in more detail below.

%%%%%%%%%%%%%%%%%%%%%%%%%%%%%%%%%%%%%%%%%%%%%%%%%%%%%%%%%%%%%%%%
%%%%%%%%%%%%%%%%%%%%%%%%%%%%%%%%%%%%%%%%%%%%%%%%%%%%%%%%%%%%%%%%

Let us consider the stationary states of the model, for which we define the notations $\sigma_P=\sigma_P(t\to\infty)$ and $\sigma_O=\sigma_O(t\to\infty)$. In these stationary states we have $d\sigma_P(t)/dt=d\sigma_O(t)/dt=0$. Taking Eq. \eqref{eq7} in the $t\to\infty$ limit, we have two solutions. One of them is $\sigma_O=0$, that represents the absorbing state where the radicalization disappears of the population. The other solution is given by a second-order polynomial for $\sigma_P$ of the form $\beta_1\sigma_P^{2} + \beta_0\sigma_P - \alpha\sigma_I = 0$. Considering the solution of such second-order polynomial and the normalization condition, Eq. \eqref{eq5}, the order parameter for the phase transition, namely the stationary fraction of opponents $\sigma_O$, is given by
\begin{equation}\label{eq9}
\sigma_O = 1 - \sigma_I - \frac{\beta_0}{2\beta_1}\{-1 \pm \sqrt{\delta}\} ~,
\end{equation}
where $\delta=1+(4\alpha\sigma_I\beta_1/\beta_0^{2})$. Numerically, we observed that the solution of Eq. \eqref{eq9} with the plus signal is the physically acceptable one, leading to $0<\sigma_O<1$.

The critical points can be found taking $\sigma_O=0$ in Eq. \eqref{eq9}. For this case, we found
\begin{equation}\label{eq10}
{\beta_{0}}_{c} = \alpha\frac{\sigma_I}{1-\sigma_I} - (1-\sigma_I)\beta_1 ~.
\end{equation}

%%%%%%%%%%%%%%%%%%%%%%%%%%%%%%%%%%%%%%%%%%%%%%%%%%%%%%%%%%%%%%%%
\begin{figure}[t]
\begin{center}
\vspace{6mm}
\includegraphics[width=0.45\textwidth,angle=0]{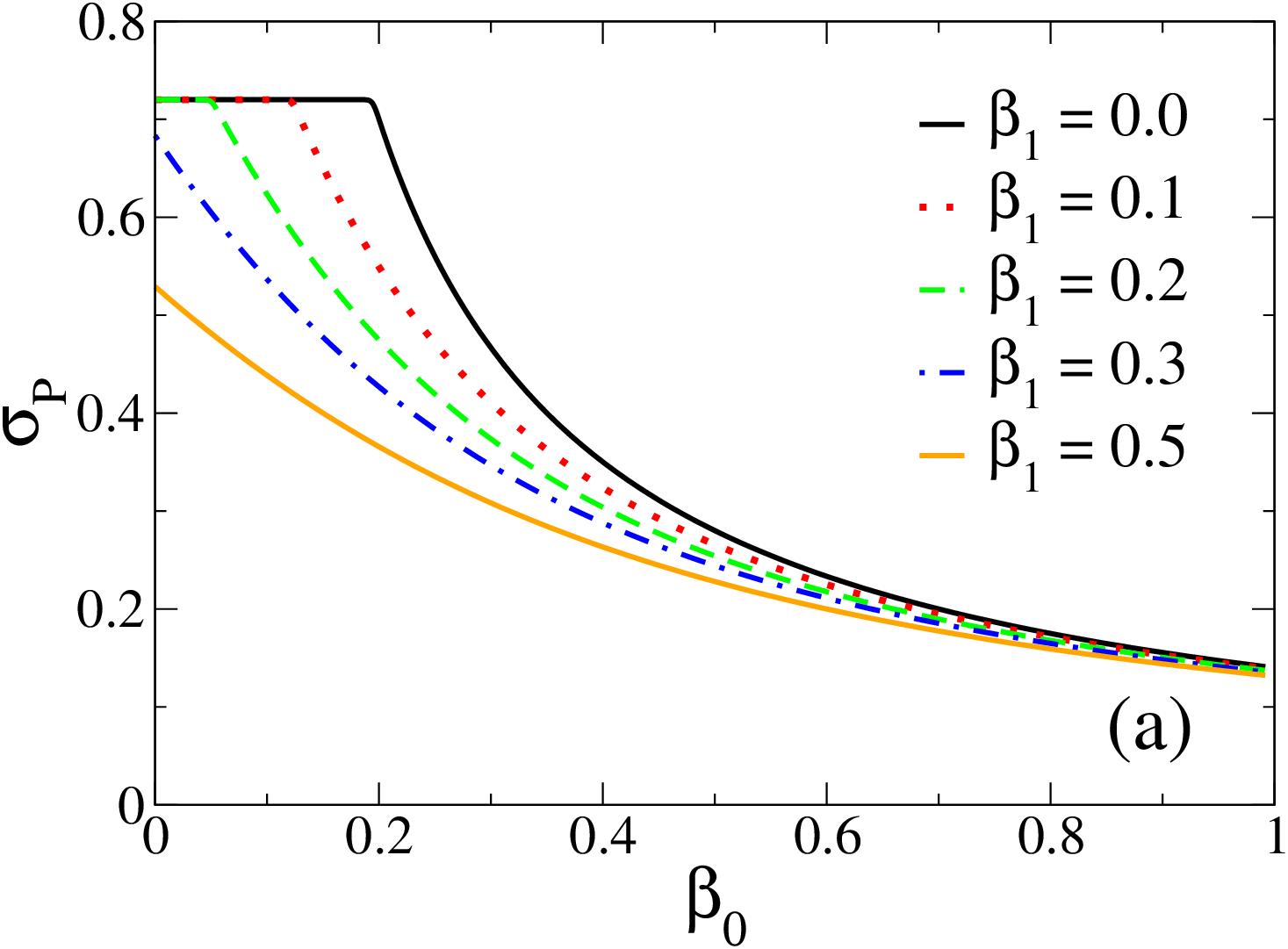}
\includegraphics[width=0.45\textwidth,angle=0]{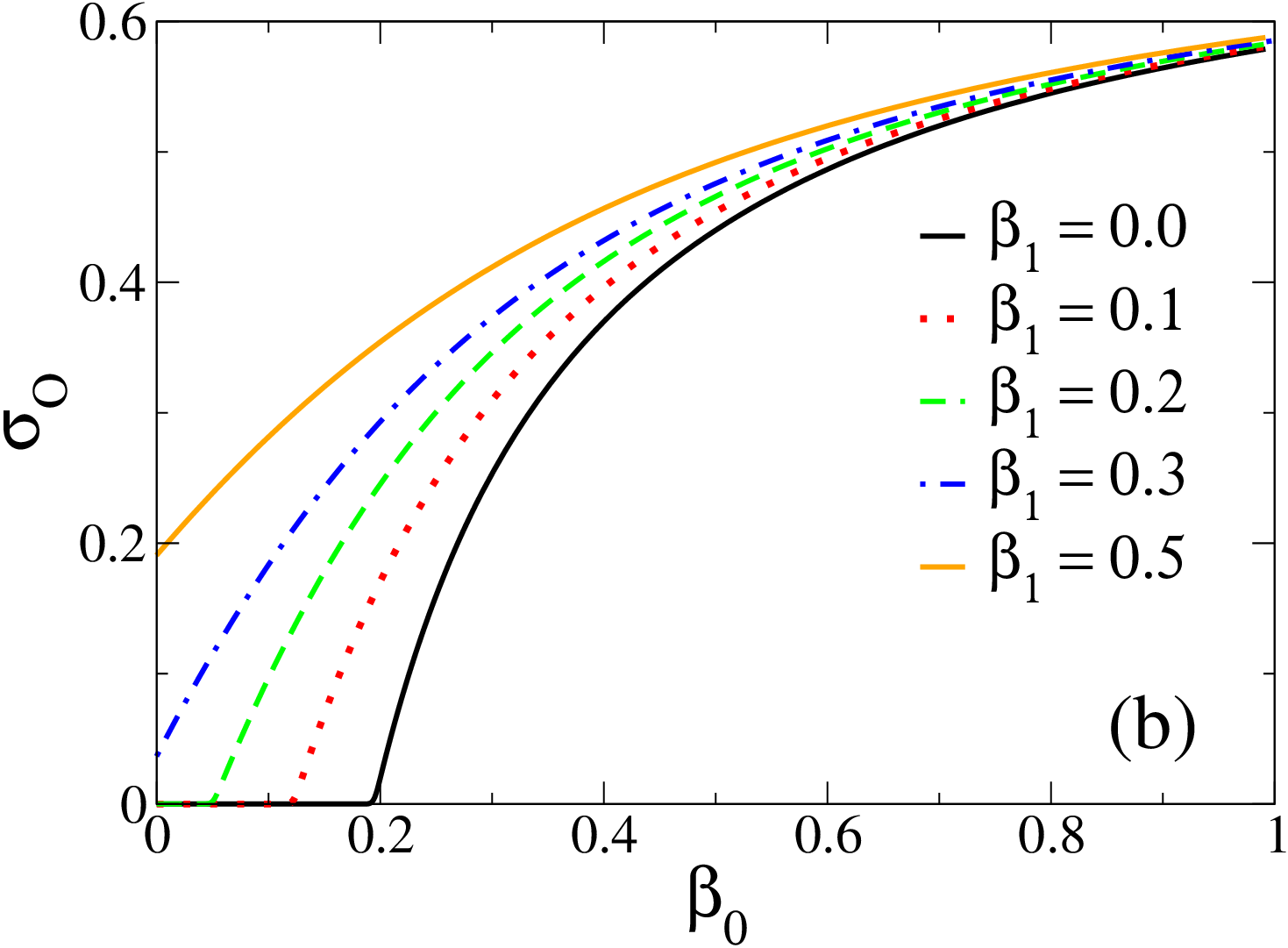}
\end{center}
\caption{(Color online) Stationary densities of peaceful agents $\sigma_P$ (panel (a)) and opponents $\sigma_O$ (panel (b)) as functions of $\beta_0$ for typical values of $\beta_1$ for the model with time-dependent rate $\beta=\beta(t)$. We can observe that for the larger values of $\beta_1$ like $0.3$ and $0.5$ there is no phase transition anymore. The fixed parameters are $\alpha=0.5$ and $\sigma_I=0.28$.}
\label{fig2}
\end{figure}
%%%%%%%%%%%%%%%%%%%%%%%%%%%%%%%%%%%%%%%%%%%%%%%%%%%%%%%%%%%%%%%%

These critical points separate the active and absorbing collective macroscopic phases, as in the original model where $\beta=\beta_0$ constant. Notice that, for $\beta_1=0$ we recover the result ${\beta_0}_c=\alpha\sigma_I/(1-\sigma_I)$ found in \cite{javarone_galam}. Now, for the time-dependet rate considering in \eqref{eq6}, the critical points depend on $\beta_1$, and the values of such critical points decrease with increasing values of $\beta_1$. In such a case, we have a limit of occurrence of the active-absorbing phase transition. The threshold value $\beta_1^{*}$ can be found taking ${\beta_0}_c=0$ in Eq. \eqref{eq10}, for which we have
\begin{equation}\label{eq11}
\beta_1^{*} = \alpha\frac{\sigma_I}{(1-\sigma_I)^{2}} ~.
\end{equation}
For $\beta_1>\beta_1^{*}$ there is no phase transition anymore. In other words,  if $\beta_1>\beta_1^{*}$ the opponents will survive in the long-time limit independent of the other parameters, and we cannot eliminate the radicalism in such situations. %Thus, in order to control violent radical activities, i.e., in order to keep a large value of $\beta_1^{*}$, two strategies can be considered: (i) increase of the social influence of inflexibles $\sigma_I$ over opponents $\sigma_O$ (i.e., increase $\alpha$); or (ii) increase the density of inflexibles, that will act in order to control de increase of the radical population (opponents).

%%%%%%%%%%%%%%%%%%%%%%%%%%%%%%%%%%%%%%%%%%%%%%%%%%%%%%%%%%%%%%%%
\begin{figure}[t]
\begin{center}
\vspace{6mm}
\includegraphics[width=0.6\textwidth,angle=0]{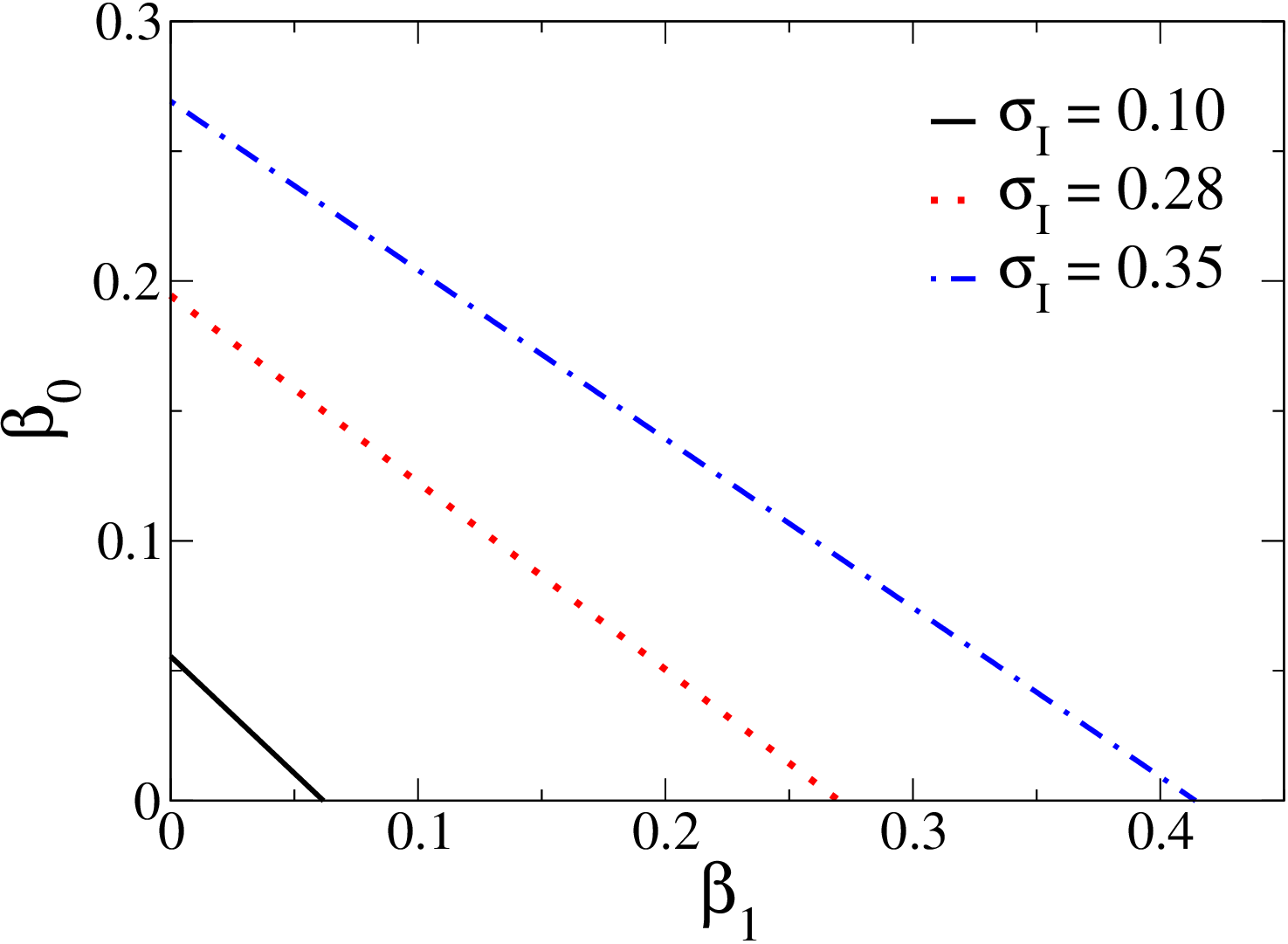}
\end{center}
\caption{(Color online) Phase diagram of the model with time-dependent rate $\beta=\beta(t)$, in the plane  $\beta_0$ versus $\beta_1$ for some values of $\sigma_I$, obtained from Eq. \eqref{eq10}. The region above the curves represent the active phase, where the opponents survive in the population in the stationary states.}
\label{fig3}
\end{figure}
%%%%%%%%%%%%%%%%%%%%%%%%%%%%%%%%%%%%%%%%%%%%%%%%%%%%%%%%%%%%%%%%

To verify those analytical results, we exhibit in Fig. \ref{fig2} the stationary densities peaceful  $\sigma_P$ and opponent agents $\sigma_O$ as functions of $\beta_0$ for typical values of $\beta_1$, since it is the novelty of the model. For such cases, we considered a typical situation analyzed in \cite{javarone_galam}, with fixed $\alpha=0.5$ and $\sigma_I=0.28$. For such values, Eq. \eqref{eq11} gives us $\beta_1^{*}\approx 0.27$. The curves were obtained by the numerical integration of Eqs. \eqref{eq7} and \eqref{eq8}. As predicted analyticaly, for $\beta_1=0.3>\beta_1^{*}$, we do not observe the absorbing state where $\sigma_O=0$ (see Fig. \ref{fig2} (b)). Considering a fixed value of $\beta_0$, we also observe that, for increasing values of $\beta_1$, the stationary value of opponents increases, and consequently the stationary value of peaceful agents decreases. Thus, the feedback mechanism adopted by opponents favors the radicalization phenomenon in the long-time limit. In other words, it is more complicated to policy makers to stop radicalization. Depending of the value of $\beta_1$, that is a feature of opponents, the radicalization cannot be eradicated (if $\beta_1>\beta_1^{*}$).

From Eq. \eqref{eq10}, let us consider a fixed value of $\alpha$. First, let us analyze the case $\beta_1=0$. For such a case, increasing the density of inflexible individuals $\sigma_I$ leads to a greater value of ${\beta_0}_c$. In other words, if the core population is large, the opponents need to increase their social pressure over the peaceful agents in order to survive in the long-time limit, i.e., they need to increase $\beta_0$. This effect is attenuated for $\beta_1>0$. Indeed, we can see that if we increase $\beta_1$, the value of the critical point ${\beta_0}_c$ decreases. In other words, even if the inflexible population increases \footnote{Of course we are talking about a population with a higher density $\sigma_I$ at the beginning, since the inflexible population is fixed.}, the social pressure $\beta(t)$ of opponents do not need to increse considerably in order to those radical individuals persist in the long run. As an illustration of such discussion, we exhibit in Fig. \ref{fig3} a phase diagram of the model in the plane $\beta_0$ versus $\beta_1$ for some values of $\sigma_I$. The region above the curves represent the active phase, where the opponents survive in the population in the stationary states.

%%%%%%%%%%%%%%%%%%%%%%%%

\subsection{Case 2: $\beta=$constant, $\alpha=\alpha(t)$}

\qquad In an analogous way we done in the previous subsection, we will consider in this section another feedback mechanism, but now in the social interaction ruled by $\alpha$. As discussed in \cite{javarone_galam}, when an inflexible agent meets an opponent it may well turn the opponent to peaceful, and for this case it was introduced a constant parameter $\alpha$. However, such exchanges could become intentional as to promote coexistence with sensitive agents via monitored informal exchanges. Thus, $\alpha$ can also be viewed as the persuasive power of inflexibles over opponent agents. In such a case, we considered that $\alpha$ is a function of time that depends of the density of opponent individuals $\sigma_O(t)$. Thus, given that the density of opponent vary with time, the inflexibles adjust their ability to convice opponents to become peaceful agents. In other words, we considered the following form for $\alpha(t)$,
\begin{equation}\label{eq12}
\alpha(t) = \alpha_0 + \alpha_1\,\sigma_O(t) ~.
\end{equation}

Thus, from Eq. \eqref{eq12} we can see that if the density $\sigma_O(t)$ increases, the inflexibles react in order to promote coexistence with sensitive agents. In this case, the inflexibles increase their ability to convince opponent agents to become peaceful ones. Thus, the system of Eqs. \eqref{eq3} and \eqref{eq4} becomes
\begin{eqnarray} \label{eq13}
\frac{d}{dt}\,\sigma_P(t) & = & \alpha_0\,\sigma_I\sigma_O(t) + \alpha_1\,\sigma_I\,\sigma_O^{2}(t) - \beta\,\sigma_O(t)\,\sigma_P(t)  \\  \label{eq14}
\frac{d}{dt}\,\sigma_O(t) & = &  \beta\,\sigma_O(t)\,\sigma_P(t) - \alpha_0\,\sigma_I\sigma_O(t) -\alpha_1\,\sigma_I\,\sigma_O^{2}(t)
\end{eqnarray}

Thus, we can see from above Eqs. \eqref{eq13} and \eqref{eq14} that the form given by Eq. \eqref{eq12} introduces another nonlinearity in the model (see the term $\sigma_O^{2}(t)$), distinct from the nonlinearity observed in the preivous subsection.

\begin{figure}[t]
\begin{center}
\vspace{6mm}
\includegraphics[width=0.4\textwidth,angle=0]{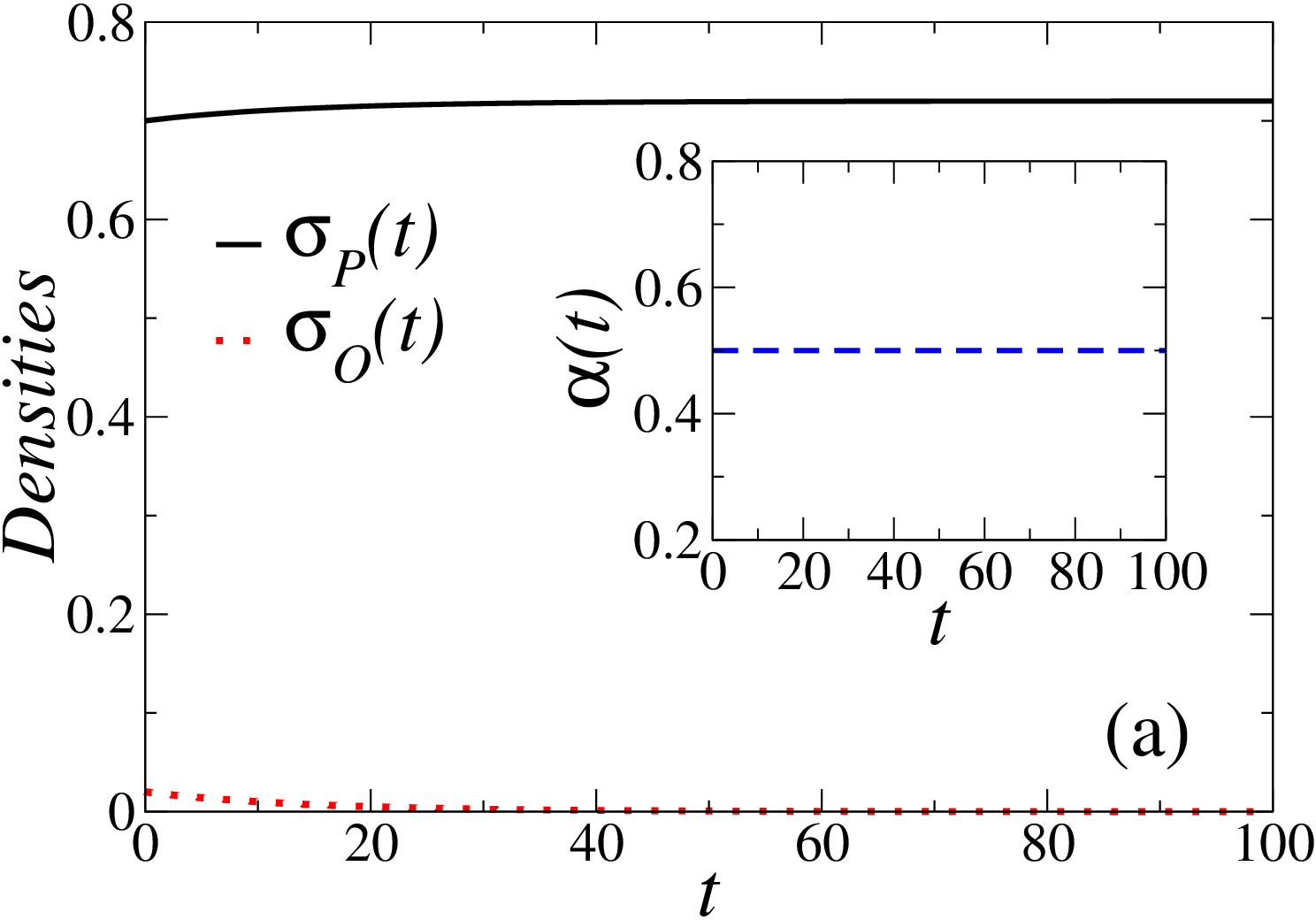}
\hspace{0.2cm}
\includegraphics[width=0.4\textwidth,angle=0]{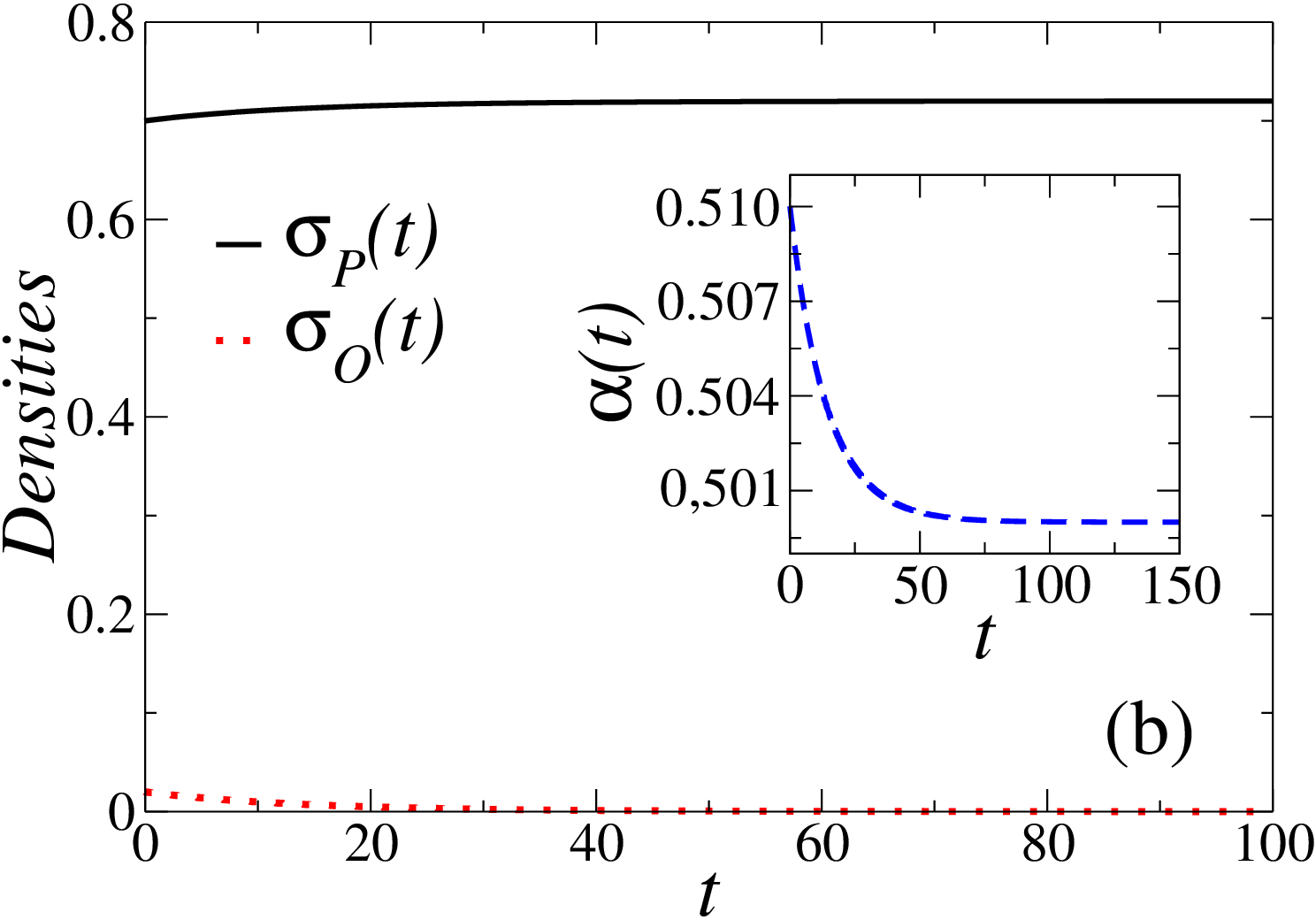}
\\
\vspace{1.0cm}
\includegraphics[width=0.4\textwidth,angle=0]{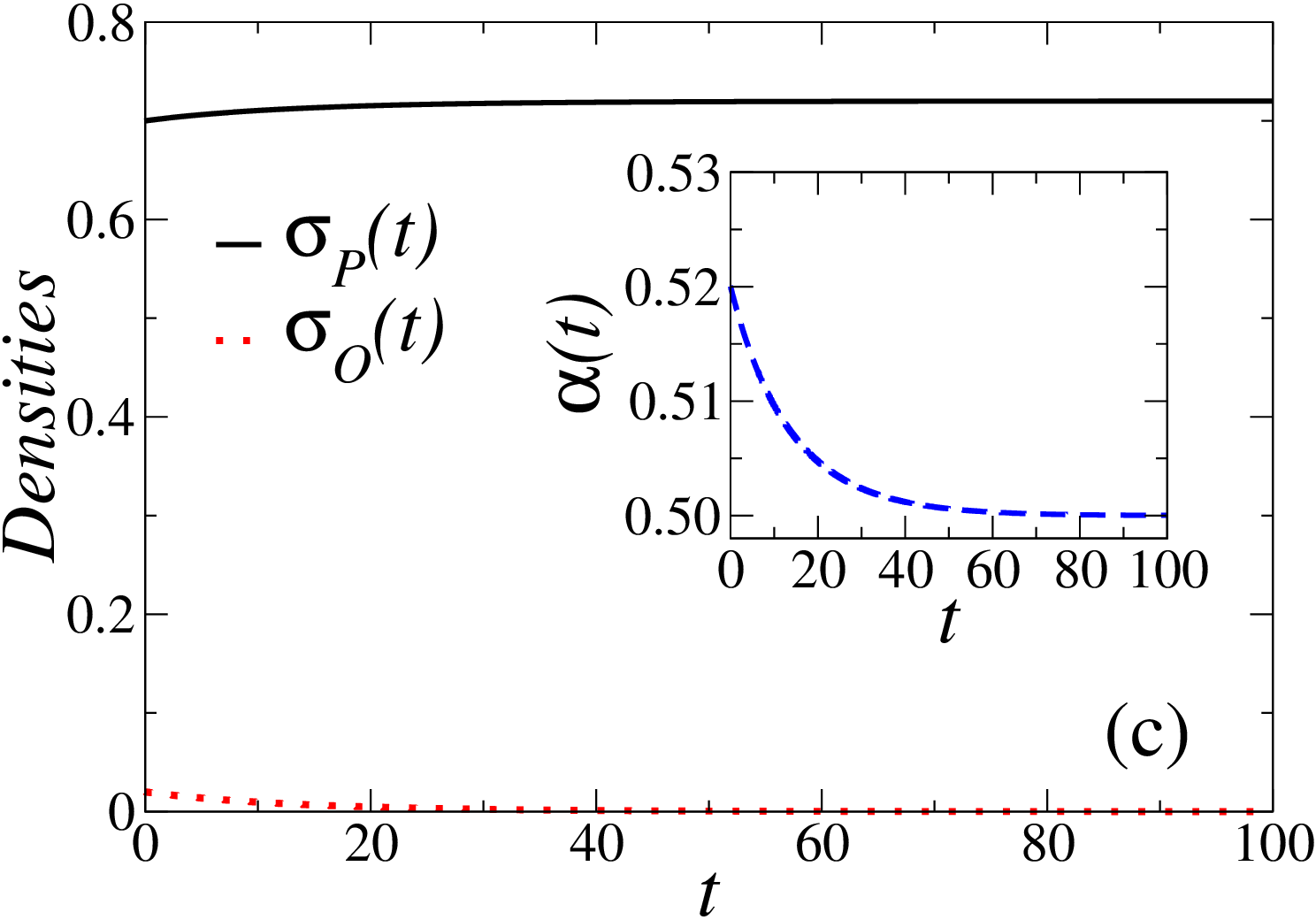}
\hspace{0.2cm}
\includegraphics[width=0.4\textwidth,angle=0]{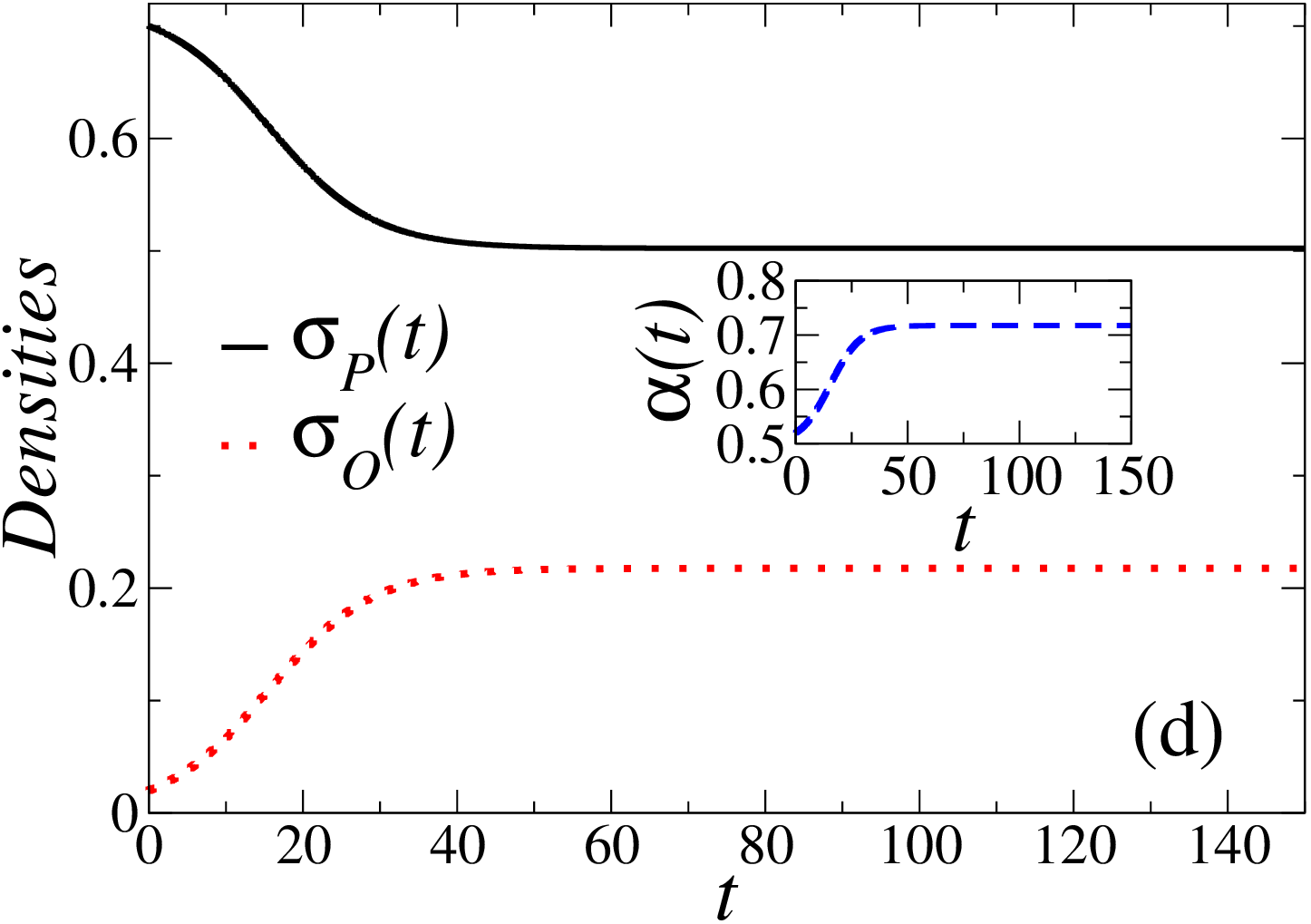}
\end{center}
\caption{(Color online) Time evolution of the densities of peaceful agents $\sigma_P(t)$ and opponents $\sigma_O(t)$ for typical values of $\alpha_1$ for the model with time-dependent rate $\alpha=\alpha(t)$: (a) $\alpha_1=0.0, \beta=0.1$ (b) $\alpha_1=0.5, \beta=0.1$ (c) $\alpha_1=1.0, \beta=0.1$; (d) $\alpha_1=1.0, \beta=0.4$. The insets exhibit the time evolution of the rate $\alpha(t)$. The fixed parameters are $\alpha_0=0.5$ and $\sigma_I=0.28$.}
\label{fig4}
\end{figure}

As in the previous subsection, we can start analyzing the time evolution of the densities $\sigma_P(t)$ and $\sigma_O(t)$. In Fig. \ref{fig4} we exhibit these quantities for fixed $\alpha_0=0.5$ and $\sigma_I=0.28$. The curves were obtained by the numerical integration of Eqs. \eqref{eq13} and \eqref{eq14}. We can see that the quantities evolve to stationary states after a long time. Differently form the previous case, subsection 3.1, for fixed and small $\beta$ like $\beta=0.1$ and increasing values of the new parameter $\alpha_1$, the absorbing state is not destroyed. As we can see in Fig. \ref{fig4}, panels (a), (b) and (c), the density of opponents evolves to $\sigma_O(t)=0$ after some time steps, even for values of $\alpha_1$ like $\alpha=0.5$ and $1.0$. However, for higher values of $\beta$ like $\beta=0.4$, the system achieves stationary states with the coexistence of peaceful agents and opponents. We will explore these states in more details in the following, through analytical calculations in the stationary states.

%%%%%%%%%%%%%%%%%%%%%%%%%%%%%%%%%%%%%%%%%%%%%%%%%%%%%%%%%%%%%%%%
%%%%%%%%%%%%%%%%%%%%%%%%%%%%%%%%%%%%%%%%%%%%%%%%%%%%%%%%%%%%%%%%

In the stationary states we have $d\sigma_P(t)/dt=0$, and Eq. \eqref{eq13} gives us two solutions. One of them is $\sigma_O=0$, that again represents the absorbing state where the radicalization disappears of the population. The other solution is given by
\begin{equation} \label{eq15}
\sigma_P = \frac{(\alpha_0+\alpha_1\sigma_O)\sigma_I}{\beta} ~.
\end{equation}
\noindent
Considering the normalization condition, Eq. \eqref{eq5}, written in the form $\sigma_P = 1-\sigma_I-\sigma_O$, Eq. \eqref{eq15} gives us
\begin{equation}\label{eq16}
\sigma_O = \frac{(1-\sigma_I)\beta - \alpha_0\sigma_I}{\alpha_1\sigma_I + \beta}
\end{equation}

The critical points can be found taking $\sigma_O=0$ in Eq. \eqref{eq16}. For this case, we found
\begin{equation}\label{eq17}
\beta_{c} = \alpha_0\frac{\sigma_I}{1-\sigma_I} ~.
\end{equation}

Notice that this result does not depend on $\alpha_1$. Indeed, it is the same result found in \cite{javarone_galam}. Thus, the time-dependence investigated here for $\alpha$ does not affect the critical behavior and the critical point of the original model. This is in contrast with we observed in the previous subsection. However, we should observe an important difference between the two formulations of the model. Whereas for the previous case the increase of the rate $\beta(t)$ leads to the increase of the opponents, which impacts on the dynamics and on the critical phenomena of the model, for the present case the increase of $\alpha$ leads to the increase of the persuasive power of inflexibles. However, the population of inflexibles is fixed, it does not vary with time. This important difference leads to the same critical point for the present case and the case $\alpha(t)=\alpha=$ constant.

\begin{figure}[t]
\begin{center}
\vspace{6mm}
\includegraphics[width=0.45\textwidth,angle=0]{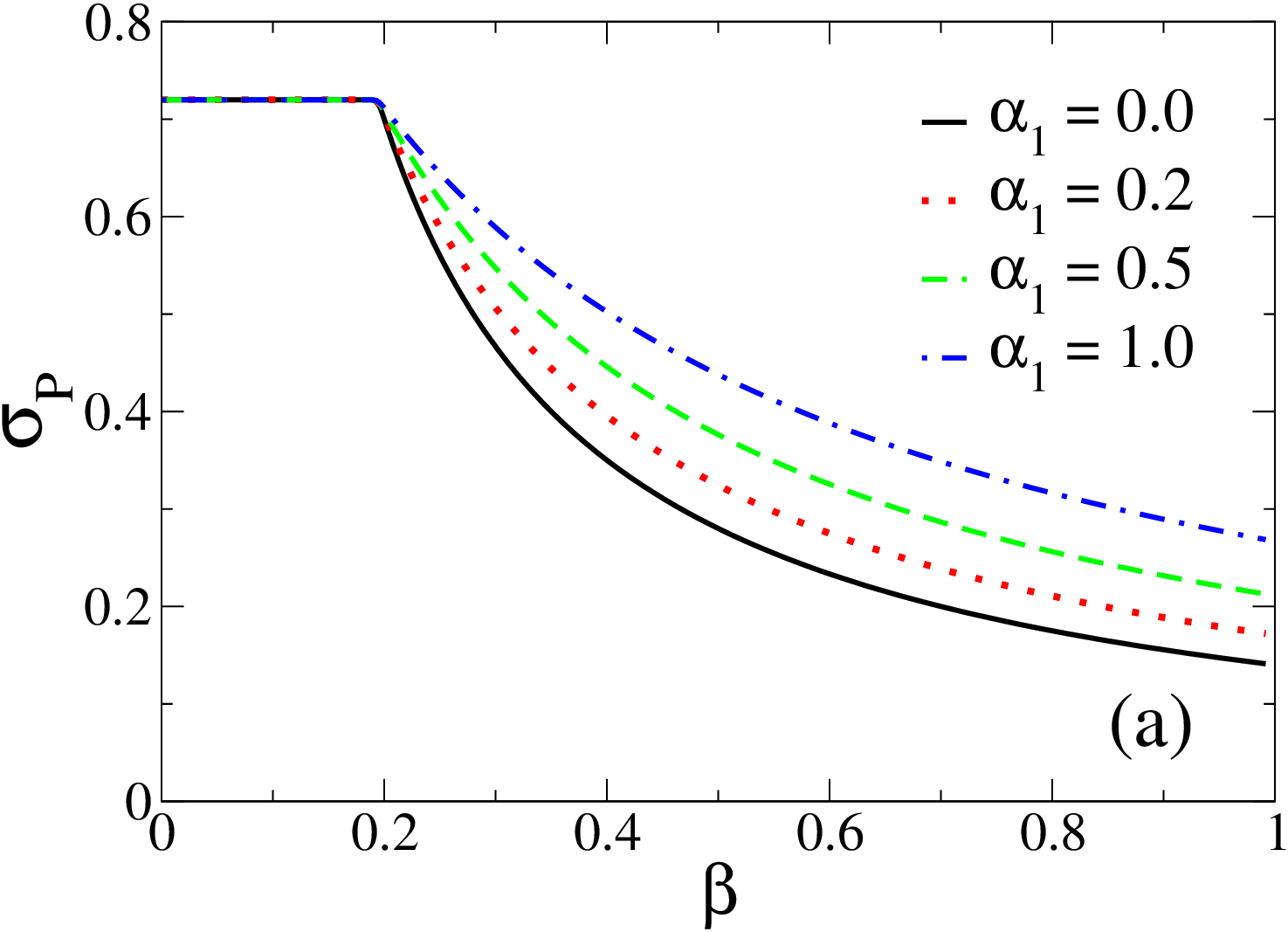}
\includegraphics[width=0.45\textwidth,angle=0]{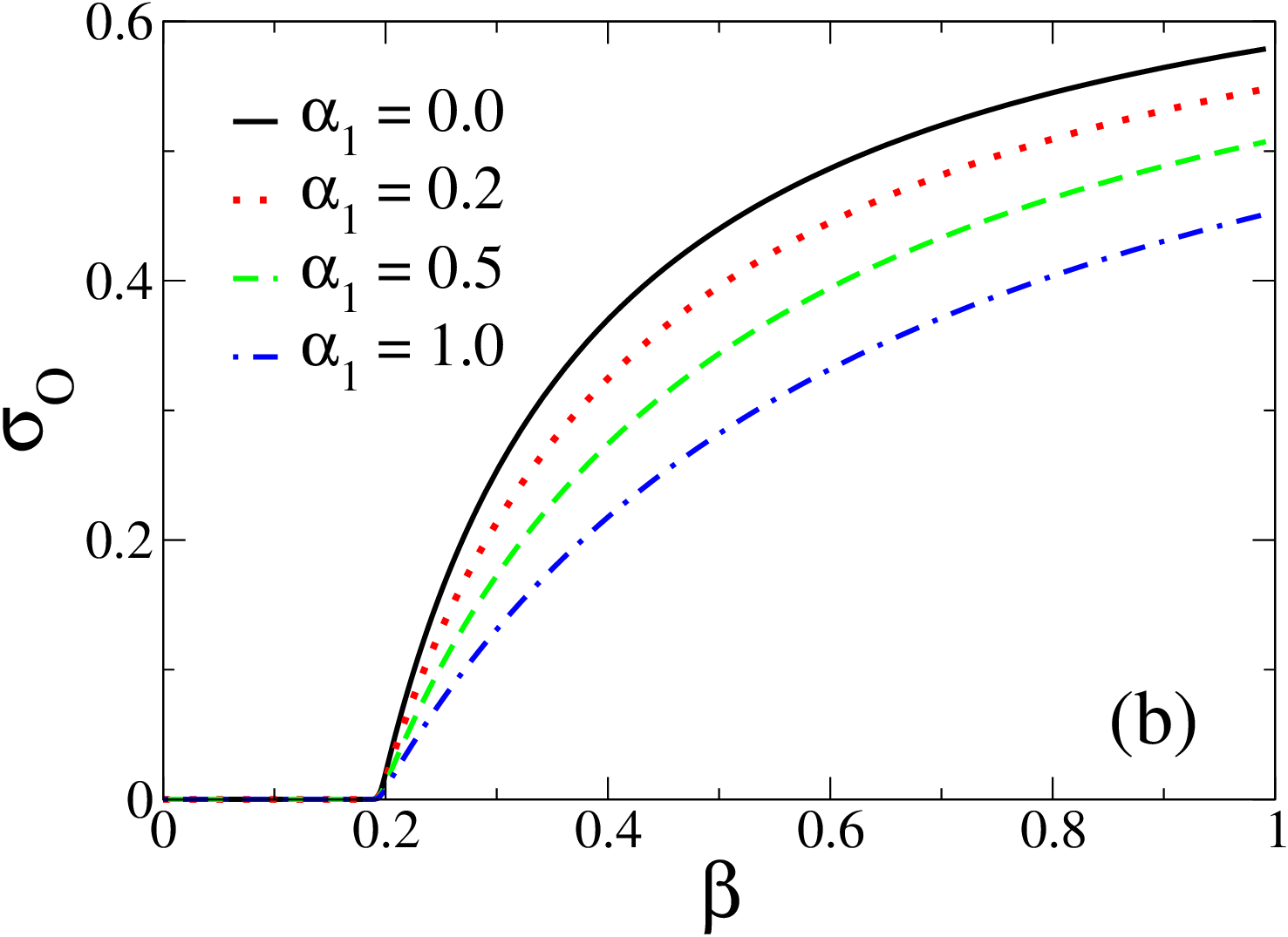}
\end{center}
\caption{(Color online) Stationary densities of peaceful agents $\sigma_P$ (panel (a)) and opponents $\sigma_O$ (panel (b)) as functions of $\beta$ for typical values of $\alpha_1$ for the model with time-dependent rate $\alpha=\alpha(t)$. We can observe that the critical point $\beta_c$ does not depend on $\alpha_1$. The fixed parameters are $\alpha_0=0.5$ and $\sigma_I=0.28$.}
\label{fig5}
\end{figure}

To verify those analytical results, we exhibit in Fig. \ref{fig5} the stationary densities peaceful  $\sigma_P$ and opponent agents $\sigma_O$ as functions of $\beta$ for typical values of $\alpha_1$, since it is the novelty of the model. For such cases, we considered a typical situation analyzed in \cite{javarone_galam} and in the previous subsection, with fixed $\alpha_0=0.5$ and $\sigma_I=0.28$. For such values, Eq. \eqref{eq17} gives us $\beta_c\approx 0.1944$. The curves were obtained by the numerical integration of Eqs. \eqref{eq13} and \eqref{eq14}. As predicted analyticaly, the critical points do not depend on $\alpha_1$ (see Fig. \ref{fig5} (b)). Considering a fixed value of $\beta$, we also observe that, for increasing values of $\alpha_1$, the starionary value of opponents decereases, and consequently the stationary value of peaceful agents increases. Thus, the feedback mechanism adopted by inflexibles weakens the radicalization phenomenon in the long-time limit. However, this is not sufficient to extinct the radicalization from the population. 

%%%%%%%%%%%%%%%%%%%%%%%%%%%%%%%%%%%%%%%%%%%%%%%%%%%%%%%%%%%%%%%%
%%%%%%%%%%%%%%%%%%%%%%%%%%%%%%%%%%%%%%%%%%%%%%%%%%%%%%%%%%%%%%%%

As a final comment, we also considered the more general case where $\alpha(t)=\alpha_0+\alpha_1\sigma_O(t)$ and $\beta(t) = \beta_0 + \beta_1\,\sigma_P(t)$. The results do not differ significantly from the previous cases. For example, considering a fixed value $\alpha_1\neq 0$, increasing $\beta_1$ leads to the destruction of the absorbing phase, as we observed in subsection $3.1$ for the case $\alpha_1=0$. On the other hand, considering a fixed value $\beta_1\neq 0$, increasing $\alpha_1$ leads the same curves as for $\alpha_1=0$, i.e., when the phase transition occurs, there is no shift of the critical point.

%%%%%%%%%%%%%%%%%%%%%%%%%%%%%%%%%%%%%%%%%%%%%%%%%

\section{Final Remarks}   

\qquad In this work we study a simple contagion model for the emergence and spreading of radicalization phenomena. For this purpose, we considered a heterogeneous population composed of core and sensitive subpopulations, and their ways of life may be at least partially incompatible. In such a case, if a conflict exist, core agents act as inflexible individuals about the issue. On the other hand, the sensitive agents choose between two options: live peacefully with core population, or oppose it. The social dynamics is ruled by pairwise interactions among agents, considering time-dependent transitions rates $\beta(t)$ and $\alpha(t)$.

First we considered a time-dependent rate $\beta=\beta(t)$ with $\alpha=$ constant, where $\beta(t)$ is a measure of the persuasive power of opponents over peaceful agents, i.e., it accounts for the rate of success of opponents in convincing peaceful agents to turn opponents. The mathematical form of the function $\beta(t)$ introduces a feedback mechanism. Such feedback mechanism adopted by opponents favors the radicalization phenomenon in the long-time limit. In other words, it is more complicated to policy makers to stop radicalization. Depending of the value of $\beta_1$, that is a feature of opponents, the radicalization cannot be eradicated. This occurs if $\beta_1>\beta_1^{*}$, where $\beta_1^{*}$ is a threshold value above which we cannot eliminate the opponents in the long-time limit . Thus, in order to control violent radical activities, i.e., in order to keep a large value of $\beta_1^{*}$, two strategies can be considered: (i) increase of the social influence of inflexibles $\sigma_I$ over opponents $\sigma_O$ (i.e., increase $\alpha$); or (ii) increase the density of inflexibles, that will act in order to control de increase of the radical population (opponents). It emphasizes the fact that, instead of being the sole prerogative of National Authorities, deradicalization would become a citizen matter \cite{javarone_galam}.

After that  we considered a time-dependent rate $\alpha=\alpha(t)$ with $\beta=$ constant, where $\alpha(t)$  is a measure of the persuasive power of inflexibles over opponents, i.e., denotes the rate per unit of time of encounters where opponents become peaceful agents. The mathematical form of the function $\alpha(t)$ introduces a feedback mechanism. Such feedback mechanism adopted by inflexibles weakens the radicalization phenomenon in the long-time limit. However, this is not sufficient to extinct the radicalization from the population, since the population of inflexibles is fixed ($\sigma_I$= constant). For such a case, to decrease radicalization it is necessary to increase the core population (inflexibles) before the occurrence of radicalization activities.

It could be interesting to consider the models on various lattices and networks. In addition, prevention programs can also be considered in the population \cite{santoprete2} in order to analyze the impact of such programs in the spreading of radicalism.

% ############################################################################

\section*{Acknowledgments}

The author acknowledges financial support from the Brazilian scientific funding agencies Conselho Nacional de Desenvolvimento Cient\'ifico e Tecnol\'ogico (CNPq, Grant 310893/2020-8) and Funda\c{c}\~ao de Amparo \`a Pesquisa do Estado do Rio de Janeiro (FAPERJ, Grant 203.217/2017).

\bibliographystyle{elsarticle-num-names}

\end{document}